\documentclass[pra,showpacs,subeqn,graphicx]{revtex4}
\usepackage[dvips]{graphicx}
\usepackage{subfigure}
\usepackage[usenames]{color}
\usepackage{amssymb}
\usepackage{amsmath}

\begin{document}

\title{Collisionally Induced Atomic Clock Shifts and Correlations}

\author{Y. B. Band and I. Osherov}
\affiliation{Departments of Chemistry and Electro-Optics and the Ilse
Katz Center for Nano-Science, \\
Ben-Gurion University, Beer-Sheva 84105, Israel}

\date{\today}

\begin{abstract}
We develop a formalism to incorporate exchange symmetry considerations
into the calculation of collisional frequency shifts for atomic clocks
using a density matrix formalism.  The formalism is developed for both
fermionic and bosonic atomic clocks.  Numerical results for a finite
temperature ${}^{87}$Sr ${}^1S_0$ ($F = 9/2$) atomic clock in a magic
wavelength optical lattice are presented.
\end{abstract}


\maketitle

\section{Introduction}

The recent progress in atomic clocks is largely due to improved
methods for cooling, trapping and manipulating cold atoms.
Improvements in the accuracy of atomic clocks continues unabated, and
reports of progress in fountain clocks \cite{Bize_04, Bauch_03} and
optical lattice clocks \cite{Ye_08, Ido_03} are published monthly.  An
atomic clock with one part in 10$^{16}$ accuracy has been recently
reported \cite{Swallows_10, Campbell_09, BlattYe_09}; this is orders
of magnitude better accuracy than the Cs standard atomic clocks.  But
collisional frequency shifts can significantly impact both the
stability and accuracy of atomic clocks.  These effects have been
systematically studied, e.g., for hydrogen \cite{Verhaar_87}, for Cs
and Rb microwave clocks \cite{Verhaar_93, Fertig_00, Leo_01} as well
as Sr optical clocks \cite{Ido_05, Swallows_10, Campbell_09,
BlattYe_09}.  It was conclusively demonstrated in the experiments
reported in Refs.~\cite{Campbell_09,BlattYe_09} on a fermionic
${}^{87}$Sr atomic clock that quantum statistics can play a critical
role in shaping interactions between atoms, and therefore in
determining collisional clock shifts.  When the temperature of the
fermionic atomic cloud is so low that the gas is degenerate, the Pauli
exclusion principle forces atoms in the gas that are in the same
internal state to occupy different vibrational levels, since their
spatial wave function must be antisymmetric.  These atoms cannot
$s$-wave scatter and therefore, at low temperatures, where higher
partial waves are frozen out, no collisional shift is possible.
Refs.~\cite{Campbell_09,BlattYe_09} showed that a small component of
the probe beam along the weak confining direction leads to different
Rabi frequencies for the weak confining direction modes, and the
optically induced inhomogeneity transfers some of the atoms into an
antisymmetric internal state, and this antisymmetric state can
$s$-wave scatter.  This gives rise to a small collisional shift.

The analysis of the experimental results in
Refs.~\cite{Campbell_09,BlattYe_09} was carried out assuming that the
{\em internal} states of two colliding atoms in the gas are in a pure
state [$\frac{1}{\sqrt{2}}(|\psi_{1} \rangle |\psi_{2} \rangle -
|\psi_{2} \rangle |\psi_{1} \rangle)$, as in Eq.~(\ref{Eq:10}), where
$|\psi_{1} \rangle$ and $|\psi_{2} \rangle$ can be superposition
states of the ground and excited clock states, as in
Eq.~(\ref{Eq:coherent})].  Moreover, reference~\cite{Rey_09}, which
introduces a microscopic many-body approach to treat collisional
shifts, Ref.~\cite{Gibble_09}, which reports exact calculations of
collisional frequency shifts for several fermions or bosons, and
Ref.~\cite{Yu_10}, which also calculates collisional shifts due to
interatomic interactions, all assume a pure {\em internal} state for
colliding atoms in their analysis of collisional shifts.  Here, we
generalize the theoretical treatment of exchange symmetry
considerations in collisional clock shifts to include mixed (density
matrix) internal states for two colliding atoms,
$\rho_{AB,\mathrm{int}}$ [see Eq.~(\ref{Eq:4a})].  That is, our
treatment does not assume that the colliding atoms are in a pure
internal state, but allows for mixed states to be treated.  We shall
also apply our analysis to bosonic-atom atomic clocks.  We assume that
the atomic gas is sufficiently cold that only $s$-wave collisions can
occur, with higher partial waves frozen out because of the angular
momentum barrier, but we do not assume that the gas is fully
degenerate.  If the atoms in the gas are fermionic, but are not in the
quantum-degenerate regime (i.e., is not described by a degenerate
fermi sea of fermionic atoms), a mixed state description is in general
required.  Similarly, bosonic atoms that are not in the
quantum-degenerate regime, require a mixed-state description.  We use
the mixed state (density matrix) description of Refs.~\cite{Band_10,
Fano_83} to develop a formalism to treat the correlation and exchange
symmetry considerations necessary to properly describe clock shifts
due to collisions of atoms in a mixed internal state.  We are
therefore able to determine how the fact that the gas in
Refs.~\cite{Campbell_09,BlattYe_09} is not quantum-degenerate affects
the calculation of the clock shifts.  As we shall show, having a mixed
state, rather than a pure state for the internal state of the
colliding atoms, does not strongly affect the calculated results
for the experimental conditions described in
Refs.~\cite{Campbell_09,BlattYe_09}.

The density matrix formalism we develop here could also allow for
calculation of blackbody radiation effects for atomic clock
transitions.  With the experimental demonstration of suppression of
collisional clock shifts due to strong-interaction many-body effects
\cite{mb_coll} (which we do not attempt to model here, since this
requires determining the correlated many-body state of the atoms in a
strongly interacting gas, and is out of the scope of this work),
calculation of blackbody radiation shifts becomes even more important,
since this can be the largest remain contribution to the error budget
of the atomic clock.

The outline of the paper is as follows.  In Sec.~\ref{Ex_sym} we
discuss the effects of exchange symmetry on the collisional frequency
shift in atomic clocks, Sec.~\ref{Fermi_clock} applies the concepts
developed in Sec.~\ref{Ex_sym} to a fermionic atomic clock.  In
Sec.~\ref{Sr_model} we calculate the collisional shift for a
${}^{87}$Sr atomic clock with parameters similar to those used in
Refs.~\cite{Campbell_09,BlattYe_09}.  Section~\ref{Bose_clock}
develops expressions for the clock shift of a bosonic atomic clock,
and a summary and conclusion are presented in
Sec.~\ref{Summary_conclusion}.

\section{Exchange symmetry considerations} \label{Ex_sym}

The density matrix characterizing systems of identical bosonic or
fermionic atoms must be properly symmetrized.  This can be
accomplished by applying symmetrization or antisymmetrization
operators.  For a degenerate gas of $N$ bosonic atoms, the state can
be symmetrized by applying the symmetrization operator ${\cal S}$,
whereas for fermionic atoms it must be antisymmetrized by applying the
antisymmetrization operator ${\cal A}$.  The symmetrization operator
for an $N$ particle state is ${\cal S} \equiv \sum_{P} P / N!$ and the
antisymmetrization operator is ${\cal A} \equiv \sum_{P} \delta_P \, P
/ N!$, where $P$ is a permutation operator, $P = \binom{{1} \, \, \,
{2} \, \, \, \ldots \, \, \, {N}}{{n_1} \, {n_2} \, \ldots \, {n_N}}$,
the sum is over all the $N!$ permutations of the atoms, and $\delta_P
= (-1)^P$ \cite{Messiah_2}.  For example, for two atoms (say, that are
undergoing a collision in a gas), ${\cal S} = \frac{1}{2} (1 +
P_{AB})$ and ${\cal A} = \frac{1}{2} (1 - P_{AB})$.  Thus, properly
symmetrized two-particle boson and fermion density matrices are of the
form
\begin{subequations}  \label{Eq:1}
\begin{equation} \label{Eq:2}
    \rho_{AB}^{\mathrm{sym}} = \frac{1}{2} (1 + P_{AB}) \rho_{AB}
    \frac{1}{2} (1 + P_{AB}) ~,
\end{equation}
\begin{equation} \label{Eq:3}
    \rho_{AB}^{\mathrm{antisym}} = \frac{1}{2} (1 - P_{AB}) \rho_{AB}
    \frac{1}{2} (1 - P_{AB}) ~.
\end{equation}
\end{subequations}

The density matrix $\rho_{AB}$ for two colliding atoms has both an
internal part, $\rho_{AB,\mathrm{int}}$ (for atoms in an atomic clock,
the internal part can be restricted to involve only two atomic
levels, the ground and excited levels participating in the optical
transition of the clock), and an external part,
$\rho_{AB,\mathrm{ext}}$, involving the motional degrees of freedom of
the atoms:
\begin{equation} \label{Eq:4a}
    \rho_{AB} = \rho_{AB,\mathrm{ext}} \otimes \rho_{AB,\mathrm{int}} 
    ~.
\end{equation}
If the atoms are free, or if they are confined by a harmonic
potential, the external part can be separated into a center of mass
and a relative motion part,
\begin{equation} \label{Eq:4b}
    \rho_{AB,\mathrm{ext}} = \rho_{AB,\mathrm{cm}} \otimes 
    \rho_{AB,\mathrm{rel}} ~.
\end{equation}
A symmetric density matrix for the internal degrees of freedom,
$\rho_{AB,\mathrm{int}} = \rho_{AB}^{\mathrm{sym}}$, must be
multiplied by a symmetric [antisymmetric] spatial density matrix
$\rho_{AB,\mathrm{rel}}$ in the spatial degree of freedom ${\bf
r}_{AB} = {\bf r}_A - {\bf r}_B$ for bosons [fermions], whereas an
antisymmetric internal density matrix, $\rho_{AB,\mathrm{int}} =
\rho_{AB}^{\mathrm{anti}}$, must be multiplied by an antisymmetric
[symmetric] spatial density matrix $\rho_{AB,\mathrm{rel}}$ for bosons
[fermions], so that the full density matrix has the right exchange
symmetry.  For cold gas temperatures, all but $l = 0$ partial waves
are frozen out due to the angular momentum barrier potential $V_l(r) =
\hbar^2 l(l+1)/(2\mu r^2)$, where $r = |{\bf r}_{AB}|$ is the relative
distance between the colliding atoms and $\mu$ is their reduced mass.
Since $s$-wave scattering wave functions are symmetric in the exchange
of the colliding particles, we conclude that at cold temperatures, the
internal density matrix must be antisymmetric for colliding fermionic
atoms and symmetric for bosonic atoms.

For the internal degrees of freedom of the colliding atoms, the
symmetrization operator is ${\cal S} = \frac{1}{2} (1 + P_{AB}) =
\frac{3}{4} + \frac{1}{4}{\boldsymbol \sigma}_A \cdot {\boldsymbol
\sigma}_B$ and the antisymmetrization operator is ${\cal A} =
\frac{1}{2} (1 - P_{AB}) = \frac{1}{4}(1 - {\boldsymbol \sigma}_A
\cdot {\boldsymbol \sigma}_B)$, hence, the density matrix for the
internal degrees of freedom are:
\begin{subequations}
\begin{equation} \label{Eq:9.2_level.1'}
    \rho_{AB}^{\mathrm{sym}} = \left(\! \frac{3}{4} +
    \frac{1}{4}{\boldsymbol \sigma}_A \cdot {\boldsymbol \sigma}_B
    \! \right) \rho_{AB} \left(\! \frac{3}{4} + \frac{1}{4}{\boldsymbol
    \sigma}_A \cdot {\boldsymbol \sigma}_B \! \right) ~,
\end{equation}
\begin{equation} \label{Eq:9.2_level.2'}
    \rho_{AB}^{\mathrm{anti}} = \frac{1}{4} \left(\! 1 -
    {\boldsymbol \sigma}_A \cdot {\boldsymbol \sigma}_B
    \! \right) \rho_{AB} \, \frac{1}{4} \left(\! 1 - {\boldsymbol
    \sigma}_A \cdot {\boldsymbol \sigma}_B \! \right) ~.
\end{equation}
\end{subequations}

Quite generally, the density matrix for the internal degrees of
freedom of two two-level systems can be written in terms of the Pauli
matrices, ${\boldsymbol \sigma}$, for each of the two-level systems as
follows \cite{Band_10,Fano_83}:
\begin{equation} \label{Eq:4}
    \rho_{AB} = \frac{1}{4} \, \left[(1 + {\bf n}_A \cdot {\boldsymbol
    \sigma}_A) \, (1 + {\bf n}_B \cdot {\boldsymbol \sigma}_B) +
    {\boldsymbol \sigma}_A \cdot {\bf C}_{AB} \cdot {\boldsymbol
    \sigma}_B \right] ~.
\end{equation}
Here ${\bf n}_i = \langle {\boldsymbol \sigma}_i \rangle =
{\mathrm{Tr}} \, \rho {\boldsymbol \sigma}_i$ is the expectation value
of the `spin' (recall that this is the spin analogy used to describe
our two-level system), as can be verified by multiplying (\ref{Eq:4})
by ${\boldsymbol \sigma}_i$ and tracing over the internal states of
atoms A and B, and the correlation matrix is $C_{ij,AB} \equiv \langle
\sigma_{i,A} \sigma_{j,B} \rangle - \langle \sigma_{i,A} \rangle
\langle \sigma_{i,B} \rangle$, as can be verified by multiplying
(\ref{Eq:4}) by $\sigma_{i,A} \sigma_{j,B}$ and tracing.  If the two
colliding two-level atoms are in an uncorrelated (and therefore
unentangled) state, ${\bf C}_{AB} = 0$, and the density matrix
$\rho_{AB}$ reduces to a tensor product of the density matrices for
each of the internal states of the atoms, $\rho_{AB} = \rho_{A}
\otimes \rho_{B}$.

\section{Fermionic atom clock}  \label{Fermi_clock}

The trap potential holding the atoms is taken to be a harmonic
oscillator potential (see Sec.~\ref{Sr_model}), but at short range,
the potential is the molecular potential of the atoms.  Hence, the
potential between two atoms in the trap is the molecular potential at
short range (nm scale) and at large distance ($\mu$m or mm scale) it
becomes the asymmetric harmonic potential confining the atoms.  The
external (trap) states of the atoms in the atomic clock will be taken
to be thermally populated harmonic oscillator states at large relative
internuclear distances (see Sec.~\ref{Sr_model}), but in this section
we concentrate on the internal state of the atoms.  The internal state
for fermionic atoms must be anti-symmetrized at low temperatures;
hence, that part of an arbitrary initial density matrix $\rho$ for two
atoms that $s$-wave scatters is given by ${\cal A} \rho_{AB} {\cal A}
= \varkappa \rho_{\mathrm{singlet}}$, where $\rho_{\mathrm{singlet}} =
\frac{1}{4}(1 - {\boldsymbol \sigma}_{A} \cdot {\boldsymbol
\sigma}_{B})$ is the density matrix of the singlet state and the
correlation coefficient $\varkappa = {\mathrm{Tr}} \, {\cal A}
\rho_{AB} {\cal A} = \frac{1}{4}\left[1 - ({\bf n}_{A}\cdot {\bf
n}_{B} + \sum_{i} C_{ii} ) \right]$, i.e.,
\begin{equation} \label{Eq:5}
    \varkappa \, \rho_{\mathrm{singlet}} = \left\{\frac{1}{4}\left[1 -
    ({\bf n}_{A}\cdot {\bf n}_{B} + \sum_{i} C_{ii} ) \right] \right
    \} \left\{ \frac{1}{4}(1 - {\boldsymbol \sigma}_{A} \cdot
    {\boldsymbol \sigma}_{B})\right \} ~.
\end{equation}
The quantity $\varkappa \equiv \frac{1}{4}\left[1 - ({\bf n}_{A}\cdot
{\bf n}_{B} + \sum_{i} C_{ii} ) \right]$ is the fraction of the
density matrix that is in the singlet state.  {\em Only this part of
the initial density matrix is able to $s$-wave scatter when the atoms
are fermions.} The expression in Eq.~(\ref{Eq:5}) plainly involves the
correlation of the spins, both because of the the spin-correlation
term involving the $C$ matrix and the ${\bf n}_{A}\cdot {\bf n}_{B}$
term.  The clock shift for fermionic atoms is given by the expression
\cite{Campbell_09, Verhaar_87, Verhaar_93, Leo_01, Harber_02,
Zwierlein_03}
\begin{equation}  \label{Eq:6}
    \Delta \nu = \frac{1}{T} \int_0^T dt \, \Delta \nu(t) = \frac{2
    \hbar}{m} a_{s,ge} \, \frac{1}{T} \int_0^T dt \, \varkappa(t)
    \left[n_g(t) - n_e(t)\right] ~,
\end{equation}
where $a_{s,ge}$ is the $s$-wave scattering length for collisions of
atoms in states $g$ and $e$, and the correlation coefficient
$\varkappa(t)$ is the instantaneous collisional shift parameter for
fermions, equal to the the pair correlation function $G^{(2)}({\bf
0})$ for collisions of pure internal states, but more generally is
given by
\begin{equation}  \label{Eq:7}
    \varkappa(t) = \frac{1}{4}\left[1 - \left({\bf n}_{A}(t) \cdot
    {\bf n}_{B}(t) + \sum_{i} C_{ii}(t) \right) \right]~.
\end{equation}
$n_g(t)$ and $n_e(t)$ are the densities of atoms in the ground and
excited state respectively [$n_g = \varrho {\overline{\rho_{gg}}}$,
$n_e = \varrho {\overline{\rho_{ee}}}$] and $\varrho = N/V$ is the
density of atoms in the gas.  Note that the instantaneous collisional
shift $\Delta \nu(t) = (2 \hbar/m) a_{s,ge} \varkappa(t) \left[n_g(t)
- n_e(t)\right]$, i.e., is proportional to $\varkappa(t) \left[n_g(t)
- n_e(t)\right]$.


If the two colliding atoms are in a pure internal state with the atoms
in the same superposition state, $|\psi_{1} \rangle_A = |\psi_{1}
\rangle_B$, then the two-particle internal state is a product state
with ${\bf n}_{A} \cdot {\bf n}_{B} = 1$ and ${\bf C}_{AB} = 0$, so
$\varkappa(t) \equiv 0$, and no collisional shift is present.  Since,
as explained in Refs.~\cite{Campbell_09,BlattYe_09}, atoms in
different trap states Rabi-flop at different rates, collisional shifts
in pure-internal states are nonvanishing.  It is instructive to
consider the case of a two-atom density matrix that corresponds to a
product state,
\begin{equation}  \label{Eq:Product_state}
    |\Psi \rangle = |\psi_{1} \rangle_A |\psi_{2} \rangle_B ~,
\end{equation}
with $\psi_{1} \ne \psi_{2}$.  This is an uncorrelated (unentangled)
pure internal state with
\begin{equation}  \label{Eq:coherent}
    |\psi_{1} \rangle_A = \alpha |g \rangle + \beta |e \rangle =
    \left( \!  \!  \begin{array}{c} \beta \\ \alpha \end{array} \!  \!
    \right) ~, \quad |\psi_{2} \rangle_B = \gamma |g \rangle + \delta
    |e \rangle = \left( \!  \!  \begin{array}{c} \delta \\ \gamma
    \end{array} \!  \!  \right) ~,
\end{equation}
whose density matrix in (\ref{Eq:4}) is also a product density density
matrix, $\rho_{AB} = \rho_{A} \otimes \rho_{B}$, and ${\bf C}_{AB} =
0$.  Hence, Eq.~(\ref{Eq:7}) becomes
\begin{equation}  \label{Eq:8}
    \varkappa(t) = \frac{1}{4}\left(1 - {\bf n}_{A}(t)\cdot {\bf
    n}_{B}(t) \right) ~,
\end{equation}
where the unit vector ${\bf n}_j$ for particle $j$ $(= A,B)$ is ${\bf
n}_j = (\sin \theta_j \cos \phi_j, \sin \theta_j \sin \phi_j, \cos
\theta_j) = (x_j,y_j,z_j)$.  Up to a multiplicative phase factor,
\begin{equation} \label{Eq:9}
    |\psi_{1} \rangle_A = \left( \!  \!  \begin{array}{c} \beta \\
    \alpha \end{array} \!  \!  \right) = \left( \!  \!
    \begin{array}{c} e^{-i \phi_A} \cos(\theta_A/2) \\ \sin(\theta_A/2)
    \end{array} \!  \!  \right)
    = 2^{-1/2} \left( \!  \!  \begin{array}{c}  \frac{x_A - i
    y_A} {\sqrt{1 - z_A}} \\ \sqrt{1 - z_A} \end{array} \!  \!
    \right) ~,
\end{equation}
and similarly for $|\psi_{2} \rangle_B = \binom{\delta}{\gamma}$.  A
simple example may be helpful.  Suppose particle $A$ is in state $|g
\rangle \equiv |\!\!\downarrow \rangle$ and particle $B$ is in state
$|e \rangle = |\!\!\uparrow \rangle$, and the particles are
uncorrelated so $\rho_{AB} = \rho_A \otimes \rho_B$; then ${\bf n}_{A}
= -\hat z$ and ${\bf n}_{B} = \hat z$, and using (\ref{Eq:8}),
$\varkappa = 1/2$.  This result can be easily understood by noting
that the state can be written as a fifty-fifty incoherent
superposition of a singlet and triplet density matrix, hence
$\varkappa = 1/2$.

Let us compare with the case when the gas is fully coherent and the
internal state of two fermions that interact via $s$-wave scattering
is in an antisymmetric pure entangled state,
\begin{equation} \label{Eq:10}
    |\Psi_{AS} \rangle = \frac{1}{\sqrt{2}}(|\psi_{1} \rangle
    |\psi_{2} \rangle - |\psi_{2} \rangle |\psi_{1} \rangle) ~.
\end{equation} 
This is the form of the internal state taken in the collisional shift
calculation of Ref.~\cite{Campbell_09}; the internal state of the two
colliding particles is a fully coherent antisymmetric state
(\ref{Eq:10}), and therefore the internal state density matrix
(\ref{Eq:4}) corresponding to this state has only one non-zero
eigenvalue, which is unity.  The collisional clock shift $\Delta \nu$
is then proportional to
\begin{equation} \label{Eq:11}
    \tilde \varkappa(t) \equiv G^{(2)}_{AS}({\bf 0}) = \langle
    \Psi_{AS} | \Psi_{AS} \rangle = 1 - \left|\alpha(t) \gamma^*(t) +
    \beta(t) \delta^*(t) \right|^2 ~.
\end{equation}
$\tilde \varkappa(t)$ for the pure state fermion case reduces to the
general result $\varkappa(t)$ in Eq.~(\ref{Eq:7}), i.e., for a pure
state, $\varkappa = \tilde \varkappa = G^{(2)}_{AS}({\bf 0})$, where
$G^{(2)}_{AS}({\bf 0})$ is the pair correlation function for fermions
at zero relative distance.  For the specific simple example suggested
above, $\alpha = 1$, $\beta = 0$, $\gamma = 0$ and $\delta = 1$, so
$\tilde \varkappa = 1$, whereas $\varkappa = 1/2$.  Hence, the
collisional clock shift for the mixed state is half that obtained for
the antisymmetric state.  Reference~\cite{Campbell_09} assumes the
internal states of the atoms in the gas is fully coherent, as in a
fully degenerate gas, and therefore the internal quantum state of the
two particles colliding in an $s$-wave collision must be
antisymmetrized as in Eq.~(\ref{Eq:10}).  We do {\em not} necessarily
assume that the gas is degenerate, hence we use a density matrix
description.  In our simple example, the two colliding particles in
the gas are either (with probability 50\%) in an internal
antisymmetric state (in which case the motional state must be
symmetric) or in an internal symmetric state (in which case the
motional state must be antisymmetric so only odd partial waves occur,
but since the gas is cold, these collisions are frozen out, i.e., only
$s$-wave collisions occur).  Hence, in the simple example considered
here, both possibilities occur with equal probability, and the factor
$\varkappa = 1/2$.  Note that, if the gas really is fully degenerate,
only the antisymmetric internal state (i.e., the singlet) is
populated, and $\varkappa$ is be unity, as obtained in
Ref.~\cite{Campbell_09}.

The collisional shift parameter $\varkappa$ for the state in the
specific simple example above is the largest possible for any product
state, and it has $\varkappa = 1/2$.  The largest possible collisional
shift parameter for an antisymmetric pure state is $\tilde \varkappa =
1$.  The smallest collisional shift is for the product state with
${\bf n}_{A}\cdot {\bf n}_{B} = 1$, i.e., $|\psi_{1} \rangle =
|\psi_{2} \rangle$, for which $\varkappa = 0$.  The smallest possible
collisional shift parameter for an antisymmetric pure state is $\tilde
\varkappa = 0$.  For any given $|\psi_{1} \rangle$ and $|\psi_{2}
\rangle$, $\varkappa$ for the product state is always smaller than
$\tilde \varkappa$ for the antisymmetric state (\ref{Eq:10}),
$\varkappa \le \tilde \varkappa$.  This suggests that it may not be
worth going to the degenerate gas limit for a fermionic gas, i.e., it
would suggest that the collisional shift is smaller if the gas is not
fully quantum degenerate.

Note that the uncorrelated case, $\rho_{AB} = \rho_{A} \otimes
\rho_{B}$ in (\ref{Eq:Product_state}), and the pure entangled state
case corresponding to $|\Psi_{AS} \rangle$ in Eq.~(\ref{Eq:10}), are
the two extreme cases of states that are possible; mixed correlated
and mixed entangled states fall in between these two limits
\cite{Band_10}.  Note also that our formulation allows us to specify
any many-body state of the gas and thereby calculate the clock shift
for arbitrary many-body gas states by tracing out all but two
particles, thereby obtaining a two-particle density matrix.

\section{Modeling the ${}^{87}$Sr optical clock}
\label{Sr_model}

In the experiments reported in Refs.~\cite{Campbell_09,BlattYe_09}, a
gas of ${}^{87}$Sr ${}^1S_0$ ($F = 9/2$) atoms in a magic wavelength
optical lattice \cite{Ye_08} with wavelength $\lambda = 813.43$ nm,
which to a good approximation forms a harmonic trap with frequencies
$\omega_x = \omega_y = 2 \pi \times 450$ Hz and $\omega_z \approx 2
\pi \times 80$ KHz, is optically pumped to the $m_F = 9/2$ (or the
$m_F = -9/2$) state.  The excited clock state, $|e \rangle (\equiv |\!
\!  \uparrow \rangle)$, is the $^3$P$_0$ state and the clock
transition has a wavelength of $\approx 698$ nm.  The gas is at a
sufficiently high temperature (e.g., $T$ = 1 $\mu$K or 3 $\mu$K) that
the harmonic oscillator motional states ($n_x$, $n_y$, $n_z$) of the
atoms in the gas can be considered populated according to a
Maxwell-Boltzmann distribution (at $T$ = 1 $\mu$K, ${\overline{n_x}} =
{\overline{n_y}} = 46)$.  The laser probing the clock transition
propagates along the 1D lattice axis, and the atoms in the gas have a
slightly different Rabi frequency $\Omega_{n_x,n_y,n_z}$ depending on
which trap state they populate.  With perfect alignment of the probe
laser along the strong confinement axis, a residual angular spread
$\Delta \theta$ between the probe laser beam and the optical lattice
${\bf k}$ remains due to the finite size of the lattice beam.
Moreover, if the symmetry is broken due to either aberrations in the
beam profile or angular misalignment $\Delta \theta$ between the
lattice and the probe beam, an even larger effective misalignment
results.  As explained in Ref.~\cite{BlattYe_09}, since the transverse
trap is isotropic, a small net misalignment angle $\Delta \theta$
along $\hat{{\bf x}}$ results in a wavevector of the probe laser that
is given by ${\bf k} \approx k(\hat{{\bf z}}+ \Delta \theta \,
\hat{{\bf x}}$).  Hence, there are slightly different Rabi frequencies
for the different trap states, $\Omega_{n_x,n_y,n_z} = \Omega_0 \,
e^{(-\eta_x^2 - \eta_y^2 - \eta_z^2)/2} L_{n_x}(\eta_x^2)
L_{n_y}(\eta_y^2) L_{n_z}(\eta_z^2)$, where $\Omega_0$ is the bare
Rabi frequency, $L_{n_{i}}$ are Laguerre polynomials, and $\eta_{i}$
are the Lamb-Dicke parameters for the transverse and longitudinal
directions, $\eta_x = \eta_y = \frac{\sin(\Delta\theta)} {\sqrt{2} \,
\lambda_L} \sqrt{\frac{\hbar}{2 m\omega_x}}$, $\eta_z =
\frac{1}{\lambda_L} \sqrt{\frac{\hbar}{2 m\omega_z}}$
\cite{Wineland_79, Campbell_09}.

We now calculate the clock shift for such a cold but non-degenerate
gas with a thermal population of trap states and an internal state so
that two colliding atoms are in a non-correlated product internal
state.  The internal state density matrix is given at any time $t$ by
a general two-qubit classically-correlated state that takes the form
\cite{Band_10}
\begin{equation} \label{Eq:qubit_CC}
    {\rho}^{\mathrm{CC}}(t) = \frac{1}{4} \sum_{k} P_k \, (1 + {\bf
    n}_{A,k}(t) \cdot {\boldsymbol \sigma}_A) \, \sum_{l} P_l \, (1 +
    {\bf n}_{B,l}(t) \cdot {\boldsymbol \sigma}_B) ~,
\end{equation}
where $k = n_x, n_y, n_z$, and $P_k = e^{-E_k/T}/\sum_{k'}
e^{-E_{k'}/T}$ is the Boltzmann distribution probability for state $k$
(and similarly for $P_l$).  Presumably $\{P_k\}$ are constant in time.
We used a Bolzmann distribution to calculate the probabilities for
finding atoms in the various trap states in the calculations for the
${}^{87}$Sr clock in Sec.~\ref{Clock_Calc}, since the experimental
temperatures used were sufficiently high for this to be an excellent
approximation.  At lower temperatures we would use the correct fermi
(or, in the case of bosons, bose) distribution.  The evolution of the
unit vectors ${\bf n}_{A,k}(t)$ and ${\bf n}_{B,k}(t)$, and the
population densities $n_g(t), n_e(t)$ (which are also functions of the
trap state $k$) are given by the evolution operator for the two-level
system in the presence of the probe laser, ${\cal U}(t,0) =
e^{-iH_{\mathrm{LM}}t/\hbar}$, where $H_{\mathrm{LM}}$ is the
light-matter Hamiltonian, is given by
\begin{equation} \label{Eq:U}
  {\cal U}(t,0) =\left(\!  \!  \begin{array}{cc} \cos(\frac{\Omega_g
  t}{2}) + \frac{i\Delta}{\Omega_g}\sin(\frac{\Omega_g t}{2}) &
  \frac{i\Omega}{\Omega_g} \sin(\frac{\Omega_g t}{2}) \\
  \frac{i\Omega}{\Omega_g}\sin(\frac{\Omega_g t}{2}) &
  \cos(\frac{\Omega_g t}{2}) - \frac{i\Delta}{\Omega_g}
  \sin(\frac{\Omega_g t}{2}) \!  \!  \end{array}\right) ~.
\end{equation}
We defined the generalized Rabi frequency, $\Omega_g \equiv
\sqrt{|\Omega_{n_x,n_y,n_z}|^2+\Delta^2}$, and the laser detuning
$\Delta = \omega - (E_e - E_g)/\hbar$.  For an atom that at $t=0$ is
in the ground electronic state and in trap state $n_x$, $n_y$, $n_z$,
the amplitudes $\alpha(t)$ and $\beta(t)$ of (\ref{Eq:9}) are
\begin{equation} \label{Eq:alpha}
      \alpha_{n_x,n_y,n_z}(t) = \cos(\frac{\Omega_g t}{2}) -
      \frac{i\Delta}{\Omega_g} \sin(\frac{\Omega_g t}{2}), \quad
      \beta_{n_x,n_y,n_z}(t) = \frac{i\Omega}{\Omega_g}
      \sin(\frac{\Omega_g t}{2}).
\end{equation}
The density matrix for the classically-correlated state can be written
in the form of Eq.~(\ref{Eq:4}) with Bloch vectors
\begin{equation} \label{Eq:qubit_CC_1}
  {\bf n}_{A}(t) = \sum_k P_k \, {\bf n}_{A,k}(t) ~, \quad {\bf
  n}_{B}(t) = \sum_l P_l \, {\bf n}_{B,l}(t) ~,
\end{equation}
and correlation matrix vanishes, $C_{ij} = 0$.  
The average clock-shift coefficient ${\overline \varkappa(t)}$, where
the average is over the trap states, is given by
\begin{equation} \label{Eq:qubit_kappa_av}
    {\overline \varkappa(t)} = \sum_{k,l} P_k P_l \frac{1 - {\bf
    n}_{A,k}(t) \cdot {\bf n}_{B,l}(t)}{4} = \frac{1}{4} [1 -
    \sum_{k,l} P_k P_l \cos \Theta_{k,l}(t)] ~,
\end{equation}
where $\Theta_{k,l}(t) = {\bf n}_{A,k}(t) \cdot {\bf n}_{B,l}(t)$.
Note that ${\overline \varkappa(t)}$ cannot be written as a product of
averages over the first and second particles separately since
$\Theta_{k,l}(t)$ generally does not factorize into a product of terms
depending separately on the first and second particle.
In the clock shift calculation we need to average over motional states
as follows:
\begin{equation}  \label{Eq:6'}
    {\overline{\Delta \nu}} = \frac{1}{T} \int_0^T dt \,
    {\overline{\Delta \nu(t)}} = \frac{2 \hbar}{m} a_{s,ge} \,
    \frac{1}{T} \int_0^T dt \, {\overline{\varkappa(t) \left[n_g(t) -
    n_e(t)\right]}} ~,
\end{equation}
where
\begin{equation} \label{Eq:qubit_kappa-ngne_av}
     {\overline{\varkappa(t) \left[n_g(t) - n_e(t)\right]}} =
     \frac{1}{4} \sum_{k} P_k \left[n_{g,k}(t) - n_{e,k}(t)\right]
     \left\{ \sum_{l} P_l \left[1 - {\bf n}_{A,k}(t) \cdot {\bf
     n}_{B,l}(t) \right] \right\} ~.
\end{equation}
Note that the quantity ${\overline{\varkappa(t) \left[n_g(t) -
n_e(t)\right]}}$ is proportional to the average instantaneous
frequency shift ${\overline{\Delta \nu}}(t)$, i.e.,
${\overline{\varkappa(t) \left[n_g(t) - n_e(t)\right]}} = \frac {m} {2
\hbar a_{s,ge}} {\overline{\Delta \nu}}(t)$.

We also calculate the collisional clock shift for a gas with thermal
population of trap states and an internal state so that two colliding
atoms are in an entangled internal state with collisional shift
parameter ${\overline{\tilde \varkappa}}(t)$,
\begin{equation} \label{Eq:qubit_tilde_kappa_av}
    {\overline{\tilde \varkappa}}(t) = \sum_{k,l} P_k P_l \sum_{k,l}
    P_k P_l \, [1 - \left|\alpha(t) \gamma^*(t) + \beta(t) \delta^*(t)
    \right|^2] ~.
\end{equation}
The instantaneous collisional clock shift in this case is proportional
to ${\overline{{\tilde \varkappa}(t) \left[n_g(t) - n_e(t)\right]}}$.
We are therefore able to compare and contrast this with ${\overline
\varkappa(t)}$.  Note that the average ${\overline{\tilde
\varkappa}}(t)$, can be written as an average of $\tilde \varkappa(t)$
over $n_x$, $n_y$, $n_z$ for each of the two particles $A$ and $B$,
but this average also cannot be expressed by taking averages over
$\alpha$, $\beta$, and $\gamma$ and $\delta$ separately because
$\tilde \varkappa(t)$ does not factorize into a product of terms
depending separately on the first and second particle.

\subsection{Results of ${}^{87}$Sr Atomic Clock Calculations}
\label{Clock_Calc}

We now present numerical results that highlight the difference between
using the two extreme limits for describing the internal states of two
colliding atoms in the gas: a product state given by $\rho_{AB} =
\rho_{A} \otimes \rho_{B}$ and a pure antisymmetric state as given in
Eq.~(\ref{Eq:10}).

\begin{figure}[!ht]
\centering
\includegraphics[width=0.45 \textwidth]{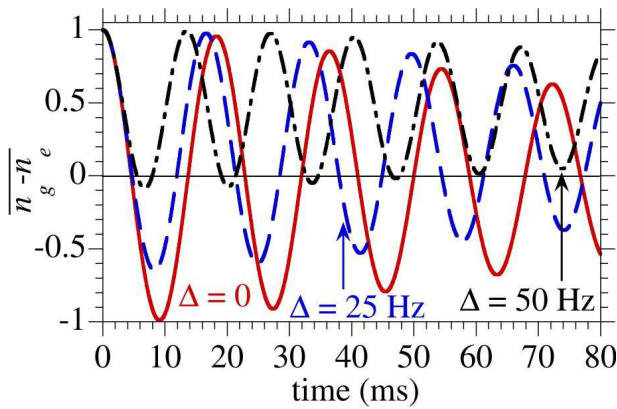}
\caption{Average population inversion, ${\overline{\left[n_g(t) -
n_e(t)\right]}}$, versus time for three different detunings for $T = 1
\, \mu$K. The fact that the Rabi frequency depends on the motional
state causes the amplitude of the Rabi oscillations to decay with
time.}
\label{Fig_BFC.n_g-n_e}
\end{figure}

In our calculations we use parameters for the system based upon those
in Refs.~\cite{Campbell_09, BlattYe_09}.  In all of our calculations
we use a Rabi frequency $\Omega_0 = 2\pi \times 59$ Hz.  Figure
\ref{Fig_BFC.n_g-n_e} shows the average population inversion
${\overline{\left[n_g(t) - n_e(t)\right]}}$ versus time for detunings
$\Delta = 2\pi \times$ 0, 25 and 50 Hz for a gas at temperature $T = 1
\, \mu$K and for a laser misalignment of $\Delta \theta = 10$ mrad.
The atoms start off at $t = 0$ in the ground electronic state and Rabi
flop with time.  For zero detuning, the oscillation of the average
population inversion with time is about zero population inversion, but
for finite $\Delta$ the time averaged population is different from
zero, and grows with detuning since less atoms are put into the
excited state for higher detuning.  The decay of the Rabi oscillations
with time results because of the thermal distribution of the atoms and
the fact that the Rabi frequencies $\Omega_{n_x,n_y,n_z}$ depend on
the trap state populated in the thermal distribution.

\begin{figure}[!ht]
\centering
\includegraphics[width=0.45 \textwidth]{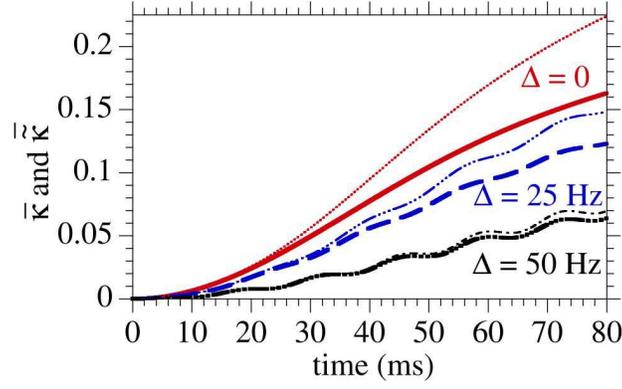}
\caption{(color online) ${\overline{\varkappa}}(t)$ (heavy curves) and
the pure-state ${\overline{\tilde \varkappa}}(t)$ (light curves)
versus time for three different detunings and for $T = 1 \, \mu$K.}
\label{Fig_BFC.kappa}
\end{figure}

Figure \ref{Fig_BFC.kappa} compares the average correlation
coefficient ${\overline{\varkappa}}(t)$ (heavy curves) and the
pure-state ${\overline{\tilde \varkappa}}(t)$ (dashed curves) versus
time for $\Delta = 2\pi \times$ 0, 25 and 50 Hz and for a gas
temperature $T = 1 \, \mu$K. The largest difference between
${\overline{\varkappa}}(t)$ and ${\overline{\tilde \varkappa}}(t)$
occurs for $\Delta = 0$, and the difference diminishes with increasing
$\Delta$.  The average correlation impacts the average collisional
frequency shift [see Eqs.~(\ref{Eq:qubit_kappa_av}), (\ref{Eq:6'}) and
(\ref{Eq:qubit_kappa-ngne_av})].  We see from the figure that these
quantities increase with the clock run-time.  For small detunings, the
average pair correlation function at zero particle separation for pure
states, ${\overline{\tilde \varkappa}}(t) = {\overline{G^{(2)}({\bf
0})}}$, is substantially larger than ${\overline{\varkappa}}(t)$ for
the product states, and the difference between these quantities
diminishes with increasing $\Delta$.  This, of course affects, the
collisional clock shift, as we shall see in the next figure.

\begin{figure}[!ht]
\centering
\includegraphics[width=0.45 \textwidth]{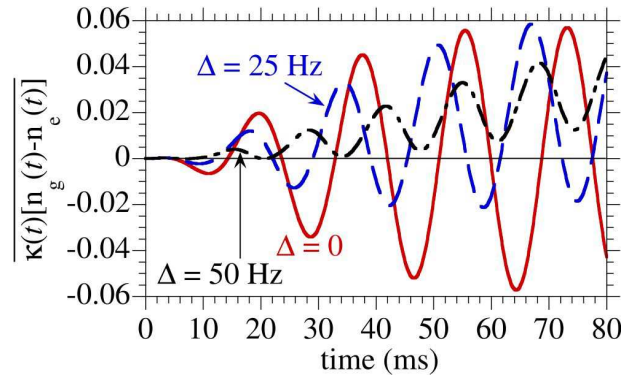}
\caption{(color online) ${\overline{\varkappa(t) \left[n_g(t) -
n_e(t)\right]}}$, which is proportional to the clock shift
${\overline{\Delta \nu(t)}}$, versus time for three different
detunings and for $T = 1 \, \mu$K.}
\label{Fig_BFC.kappa_n_g-n_e}
\end{figure}

Figure \ref{Fig_BFC.kappa_n_g-n_e} shows the quantity
${\overline{\varkappa(t) \left[n_g(t) - n_e(t)\right]}}$ versus time
calculated for three detunings, $\Delta = 2\pi \times$ 0, 25 and 50
Hz, and for $T = 1 \, \mu$K. This quantity is proportional to the
instantaneous clock shift ${\overline{\Delta \nu(t)}}$ (but since the
$s$-wave scattering length $a_{s,ge}$ is not well known, the factor
$(2 \hbar/m) a_{s,ge}$ in Eq.~(\ref{Eq:6}) is removed from the
quantity plotted).  It is clear from Fig.~\ref{Fig_BFC.kappa_n_g-n_e},
and from Eq.~(\ref{Eq:6}), that the temporal average of the clock
shift for $\Delta = 0$ will be close to zero if the temporal average
is taken over a time $t_f = T$, which corresponds to a full Rabi
cycle, but for finite detuning, the temporally averaged clock shift
grows with detuning.

\begin{figure}[!ht]
\centering
\includegraphics[width=0.45 \textwidth]{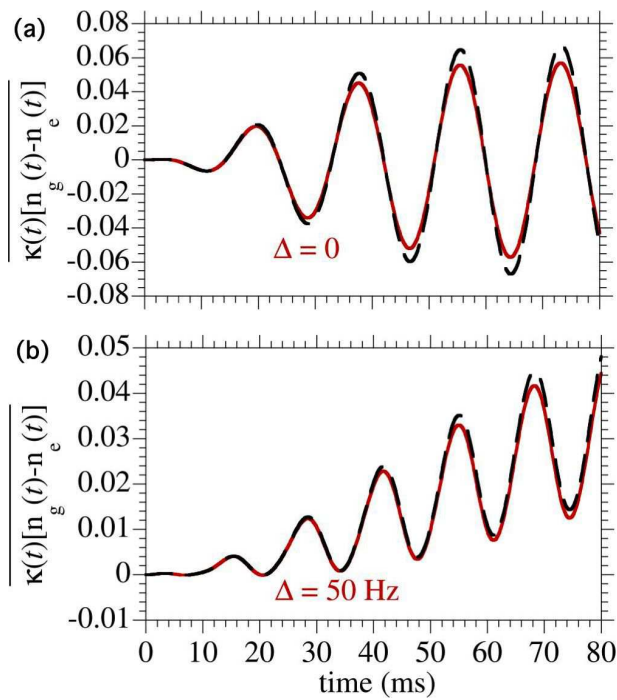}
\caption{(color online) (a) ${\overline{\varkappa(t) \left[n_g(t) -
n_e(t)\right]}}$ (solid red curves), which is proportional to the
clock shift ${\overline{\Delta \nu(t)}}$, and the pure-state
${\overline{{\tilde \varkappa}(t) \left[n_g(t) - n_e(t)\right]}}$
(dashed black curves) versus time for detuning $\Delta = 0$, and (b)
$\Delta = 2 \pi \times 50$ Hz.  The gas temperature is taken to be $T
= 1 \, \mu$K.}
\label{Fig_BFC.kappa_n_g-n_e_comp}
\end{figure}

\begin{figure}
\centering
\includegraphics[width=0.45 \textwidth]{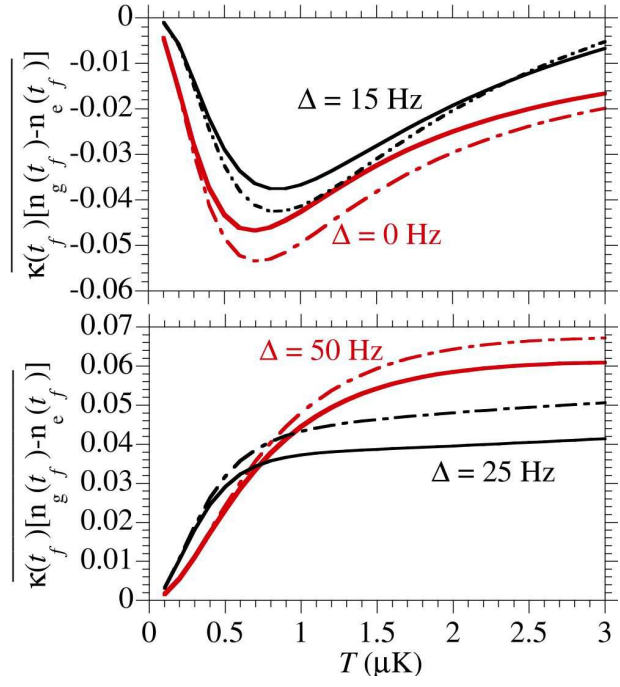}
\caption{(color online) ${\overline{\varkappa(t_f) \left[n_g(t_f) -
n_e(t_f)\right]}}$ (solid curves) and ${\overline{{\tilde
\varkappa}(t_f) \left[n_g(t_f) - n_e(t_f)\right]}}$ (dashed curves)
versus temperature for $t_f = 80$ ms and for four 4 detunings, (a)
$\Delta = 0$ and $\Delta = 2 \pi \times 15$ Hz, and (b) $\Delta = 2
\pi \times 25$ and $\Delta = 2 \pi \times 50$ Hz.}
\label{Fig_BFC.Kap_ng-ne.vs.T}
\end{figure}

Figure \ref{Fig_BFC.kappa_n_g-n_e_comp} compares the product state
${\overline{\varkappa(t) \left[n_g(t) - n_e(t)\right]}}$ (solid red
curves) and the pure-state ${\overline{{\tilde \varkappa}(t)
\left[n_g(t) - n_e(t)\right]}}$ (dashed black curves) for $\Delta = 0$
and $\Delta = 2 \pi \times 50$ Hz.  The $\approx 25$\% difference
between ${\overline{\varkappa(t)}}$ and ${\overline{{\tilde
\varkappa}(t)}}$ evident in Fig.~\ref{Fig_BFC.kappa} for $\Delta = 0$
at large times shows up mostly at the extrema of
${\overline{\varkappa(t) \left[n_g(t) - n_e(t)\right]}}$ and
${\overline{{\tilde \varkappa}(t) \left[n_g(t) - n_e(t)\right]}}$ in
Fig~.\ref{Fig_BFC.kappa_n_g-n_e_comp}; elsewhere the
population-difference $\left[n_g(t) - n_e(t)\right]$ becomes small and
the difference between the averaged quantities plotted in
Fig.~\ref{Fig_BFC.kappa_n_g-n_e_comp} become small and hard to see.
As the detuning increases, the magnitude of the collisional phase
shift is greater for ${\overline{{\tilde \varkappa}(t) \left[n_g(t) -
n_e(t)\right]}}$ than for ${\overline{\varkappa(t) \left[n_g(t) -
n_e(t)\right]}}$, but the differences in ${\overline{\varkappa(t)}}$
and ${\overline{{\tilde \varkappa}(t)}}$ decrease with increased
detuning, as evident from Fig.~\ref{Fig_BFC.kappa}.  Thus, the
non-degenerate gas has a lower clock shift (in magnitude) than the gas
degenerate with an antisymmetric internal state.  The effect appears
to be very small, but in fact, we shall see in
Fig.~\ref{Fig_BFC.Kap_ng-ne.vs.T} that, partly due strong cancellation
of the positive and negative clock shifts [particularly for $\Delta =
0$ -- see Fig.~\ref{Fig_BFC.kappa_n_g-n_e_comp}(a)], the time averaged
clock shifts ${\overline{\varkappa(t_f) \left[n_g(t_f) -
n_e(t_f)\right]}}$ and ${\overline{{\tilde \varkappa}(t_f)
\left[n_g(t_f) - n_e(t_f)\right]}}$ are not inconsequential.

\begin{figure}
\centering
\includegraphics[width=0.45 \textwidth]{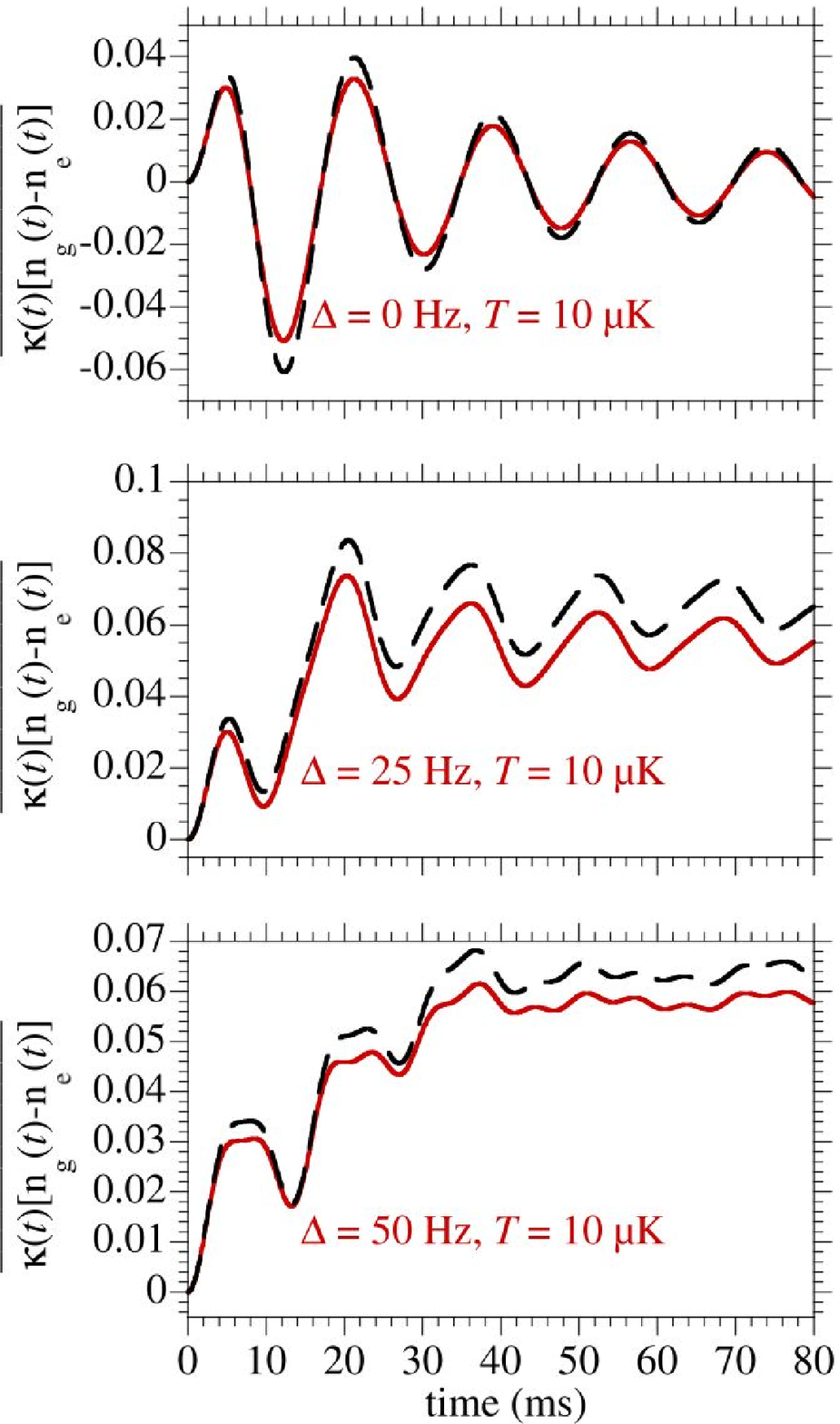}
\caption{(color online) ${\overline{\varkappa(t) \left[n_g(t) -
n_e(t)\right]}}$ (solid red curves) and ${\overline{{\tilde
\varkappa}(t) \left[n_g(t) - n_e(t)\right]}}$ (dashed black curves)
versus time for 3 detunings, $\Delta = 0$, $\Delta = 2 \pi \times 25$
and $\Delta = 2 \pi \times 50$ Hz, and for $T = 10 \, \mu$K. Note the
different ordinate scales.}
\label{Fig_BFC.Kappa_ng_ne.0-50Hz.T=10mK}
\end{figure}

Figure \ref{Fig_BFC.Kappa_ng_ne.0-50Hz.T=10mK} compares
${\overline{\varkappa(t) \left[n_g(t) - n_e(t)\right]}}$ (solid red
curves) and ${\overline{{\tilde \varkappa}(t_f) \left[n_g(t) -
n_e(t)\right]}}$ as a function of time for three different detunings
at a relatively high temperature of $T = 10 \, \mu$K. The magnitude of
the collisional phase shift is a little larger for ${\overline{{\tilde
\varkappa}(t) \left[n_g(t) - n_e(t)\right]}}$ than for
${\overline{\varkappa(t) \left[n_g(t) - n_e(t)\right]}}$.  The
difference increases with increasing detuning, but then saturates.
The behavior here for $T = 10 \, \mu$K is actually quite a bit
different from that shown in Fig.~\ref{Fig_BFC.kappa_n_g-n_e_comp} for
$T = 1 \, \mu$K. For example, the decay of the clock shift
oscillations as a function of time at $\Delta = 0$ in
Fig.~\ref{Fig_BFC.Kappa_ng_ne.0-50Hz.T=10mK} is not evident in
Fig.~\ref{Fig_BFC.kappa_n_g-n_e_comp}, and the oscillations with time
at $\Delta = 2 \pi \times$50 Hz in
Fig.~\ref{Fig_BFC.kappa_n_g-n_e_comp} are largely damped in
Fig.~\ref{Fig_BFC.Kappa_ng_ne.0-50Hz.T=10mK}.  Here (and in
Fig.~\ref{Fig_BFC.kappa_n_g-n_e_comp}), for zero detuning, the
instantaneous clock shift oscillates around zero with time but this
changes with increased detuning.  Note that the ordinate scales of the
three frames in Fig.~\ref{Fig_BFC.Kappa_ng_ne.0-50Hz.T=10mK} are
different.

\begin{figure}[!ht]
\centering
\includegraphics[width=0.45 \textwidth]{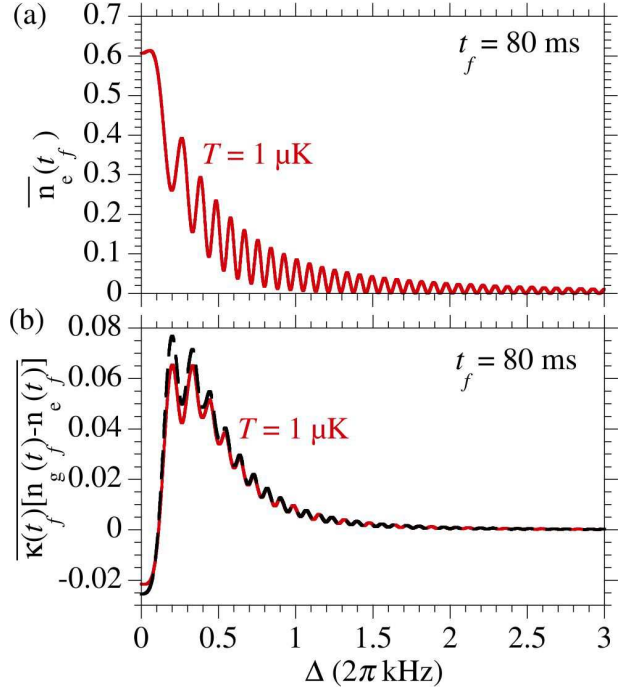}
\caption{(a) Average excited state population ${\overline{n_e(t_f)}}$
versus detuning $\Delta$ for $t_f = 80$ ms.  (b)
${\overline{\varkappa(t_f) \left[n_g(t_f) - n_e(t_f)\right]}}$ (solid
red curve) and ${\overline{{\tilde \varkappa}(t_f) \left[n_g(t_f) -
n_e(t_f)\right]}}$ (dashed black curve) versus detuning $\Delta$ for
$t_f = 80$ ms.}
\label{Fig_BFC.Kap_ng_ne_80ms_vs_D}
\end{figure}

\begin{figure}[!ht]
\centering
\includegraphics[width=0.45 \textwidth]{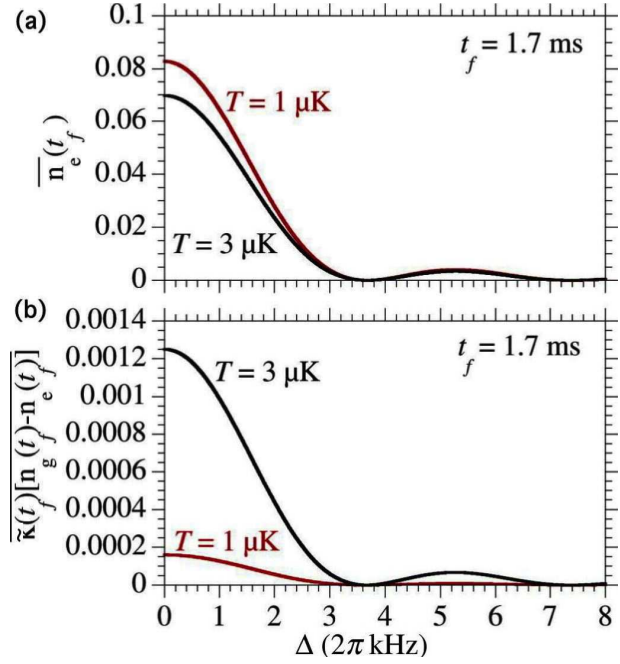}
\caption{(color online) (a) $n_e(t_f)$ and (b) ${\overline{{\tilde
\varkappa}(t_f) \left[n_g(t_f) - n_e(t_f)\right]}}$ versus detuning
$\Delta$ for $t_f = 1.7$ ms.}
\label{Fig_BFC.Kappa_ng_ne_1.7_vs_D}
\end{figure}

Figure \ref{Fig_BFC.Kap_ng_ne_80ms_vs_D}(a) plots the average excited
state population ${\overline{n_e(t_f)}}$ and
Fig.~\ref{Fig_BFC.Kap_ng_ne_80ms_vs_D}(b) the average clock shift
${\overline{{\tilde \varkappa}(t) \left[n_g(t_f) - n_e(t_f)\right]}}$
versus detuning $\Delta$ for $t_f = 80$ ms (note that for
Figs.~\ref{Fig_BFC.Kap_ng_ne_80ms_vs_D}-\ref{Fig_BFC.Kap_ng-ne.vs.T},
the bar indicating average also means average over the clock time
$t_f$).  The population for $\Delta = 0$ and $t_f = 80$ ms is about
60\% in the excited state, so ${\overline{n_e(t_f)}} >
{\overline{n_g(t_f)}}$.  This is why the quantity ${\overline{{\tilde
\varkappa}(t_f) \left[n_g(t_f) - n_e(t_f)\right]}}$ is negative for
$\Delta = 0$.  At the detuning $\Delta \approx 2 \pi \times 0.15$ Hz
for which ${\overline{n_e(t_f)}} = 0.5$, clearly, ${\overline{{\tilde
\varkappa}(t_f) \left[n_g(t_f) - n_e(t_f)\right]}} = 0$.  At higher
detunings, the oscillations in ${\overline{{\tilde \varkappa}(t_f)
\left[n_g(t_f) - n_e(t_f)\right]}}$ as a function of detuning is due
to the oscillations in the population ${\overline{n_e(t_f)}}$ versus
$\Delta$.

Figure \ref{Fig_BFC.Kappa_ng_ne_1.7_vs_D} plots
${\overline{n_e(t_f)}}$ and ${\overline{{\tilde \varkappa}(t_f)
\left[n_g(t_f) - n_e(t_f)\right]}}$ versus detuning $\Delta$ for $T =
1 \, \mu$K and $T = 3 \, \mu$K at a pulse duration of $t_f = 1.7$ ms,
which is near the start of the first Rabi cycle --- see
Fig.~\ref{Fig_BFC.n_g-n_e}.  Both the average excited state fraction
in Fig.~\ref{Fig_BFC.Kappa_ng_ne_1.7_vs_D}(a) and the collisional
shift in Fig.~\ref{Fig_BFC.Kappa_ng_ne_1.7_vs_D}(b) are smaller for
the larger temperature.  The collisional shift is seen to be very
substantially smaller at $T = 1 \, \mu$K than for $T = 3 \, \mu$K in
Fig.~\ref{Fig_BFC.Kappa_ng_ne_1.7_vs_D}(b).  The temperature effect is
dramatic for all detunings, i.e., ${\overline{{\tilde \varkappa}(t_f)
\left[n_g(t_f) - n_e(t_f)\right]}}$ is significantly reduced as the
temperature is decreased from $T = 3 \, \mu$K to $T = 1 \, \mu$K for
virtually all detunings.  At $\Delta = 0$, ${\overline{ \varkappa(t_f)
\left[n_g(t_f) - n_e(t_f)\right]}}$ is positive [in contrast with
Fig.~\ref{Fig_BFC.Kap_ng_ne_80ms_vs_D} for $t_f = 80$ ms], since this
is at the start of the first Rabi cycle (see
Fig.~\ref{Fig_BFC.n_g-n_e}) and $n_e(t_f) \ll n_g(t_f)$.  Moreover,
very little vestige remains of the oscillations with detuning that was
so prominent in Fig.~\ref{Fig_BFC.Kappa_ng_ne_1.7_vs_D}.
%
It is interesting to note that if one plots ${\overline{{\tilde
\varkappa}(t_f) \left[n_g(t_f) - n_e(t_f)\right]}}$ versus $n_e(t_f)$
[i.e., if one plots the ordinate of
Fig.~\ref{Fig_BFC.Kappa_ng_ne_1.7_vs_D}(b) versus the ordinate of
Fig.~\ref{Fig_BFC.Kappa_ng_ne_1.7_vs_D}(a)], the collisional shift is
nearly linear with $n_e(t_f)$ for both $T = 1 \, \mu$K and $T = 3 \,
\mu$K, but the slope is strongly temperature dependent and increases
with increasing temperature.  Holding $n_e(t_f)$ (or the excitation
fraction) constant, and lowering the temperature of the gas, serves to
significantly lower clock shifts in this regime of low excitation
fraction.

Figure~\ref{Fig_BFC.Kap_ng-ne.vs.T} plots the clock shift $\Delta \nu$
for $t_f = 80$ ms as a function of temperature for four different
detunings.  For $\Delta = 0$ and $\Delta = 2 \pi \times 15$ Hz, the
collisional shift first decreases with temperature from zero to around
a value of $-0.05$, and at around 0.75 $\mu$K, $\Delta \nu$ begins to
increase with temperature.  At large temperatures, $\Delta \nu$
appears to tend to zero.  For the higher detunings of $\Delta = 2 \pi
\times 25$ and $\Delta = 2 \pi \times 50$ Hz shown in the second
panel, the collisional shift increases with increasing temperature and
then saturates.  Again, the magnitude of ${\overline{\varkappa(t)
\left[n_g(t) - n_e(t)\right]}}$ is almost always smaller than
${\overline{{\tilde \varkappa}(t_f) \left[n_g(t_f) -
n_e(t_f)\right]}}$.  Moreover, we see a very different temperature
dependence of the clock shift for small detunings than for larger
detunings.

\section{Bosonic atom clock}  \label{Bose_clock}


Let us now consider a non-degenerate gas of bosonic atoms that is
nevertheless sufficiently cold so that only $s$-wave collisions
participate in the collisional frequency shift.  For example, consider
an atomic clock based on neutral bosonic atoms trapped in a deep
``magic-wavelength'' optical lattice, or more specifically, $^{88}$Sr
atoms in the 5s$^2$ $^1$S$_0$ $|F = 0,M_F= 0 \rangle$.  A weak
transition to the $^3$P$_0 \, |F = 0,M_F= 0 \rangle$ state is
available when a static magnetic field enables a direct optical
excitation of forbidden electric-dipole character, that is otherwise
prohibitively weak (the transition occurs by mixing the $^3$P$_1$ with
the $^3$P$_0$ state) \cite{Hollberg_06}.  Such an atomic clock, probed
using Ramsey fringe spectroscopy \cite{Vanier_Lemonde}, was analyzed
in Ref.~\cite{Band_00}.  For the collision of two bosons in such a gas
via $s$-wave collisions, the internal two-particle state must also be
symmetric.  The internal states of the atoms in such a collision can
be of the form $|\!\!\uparrow\downarrow \rangle +
|\!\!\downarrow\uparrow \rangle$, $|\!\downarrow \downarrow \rangle$
or $|\!  \!  \uparrow\uparrow \rangle$, where $|\!\!\downarrow \rangle
\equiv |g \rangle$ and $|\!\!\uparrow \rangle \equiv |e \rangle$.  The
$s$-wave scattering lengths for these states are denoted $a_{s,ge}$,
$a_{s,gg}$ and $a_{s,ee}$ respectively.  The Hamiltonian is given by
\cite{Band_00},
\begin{equation}  \label{Eq:H}
\hat H = \sum_{i=g,e} \Delta (a^\dag_{e}\hat a_{e} -a^\dag_{g}\hat
a_{g}) - \frac{\hbar \Omega(t)}{2} (\hat a^\dag_{g} \hat a_{e} + \hat
a^\dag_{e} \hat a_{g}) + \frac{4\pi\hbar^2}{m} \sum_{i,j=g,e} a_{s,ij}
\hat a^\dag_{i} \hat a^\dag_{j}\hat a_{j}\hat a_{i} \,.
\end{equation}
This leads to the following formula for the collisional frequency
shift for bosons \cite{Leo_01, Harber_02, Zwierlein_03}:
\begin{equation}
    \Delta \nu = \frac{2 \hbar}{m} \, \frac{1}{T} \int_0^T dt \,
    \left\{ a_{s,ge} \varkappa_{\frac{|\!\uparrow\downarrow\rangle +
    |\!\downarrow \uparrow \rangle}{\sqrt{2}}}(t) \left[n_g(t) -
    n_e(t) \right] + a_{s,gg} \varkappa_{|\!\downarrow \downarrow
    \rangle}(t) n_g(t) + a_{s,ee} \varkappa_{|\!\uparrow \uparrow
    \rangle}(t) n_e(t) \right\} ~.
\end{equation}
This expression can be compared with Eq.~(\ref{Eq:6}) for the fermion
case.  Here, clock shifts are due to $gg$ $s$-wave collisions, $ee$
$s$-wave collisions and $ge$ $s$-wave collisions.  The three
$\varkappa$ correlation parameters that multiply these scattering
lengths can be determined by calculating the projection of the density
matrix of two colliding particles onto the symmetric states,
$\frac{1}{\sqrt{2}}(|\!\!\uparrow\downarrow\rangle +
|\!\!\downarrow\uparrow\rangle)$, $|\!\!\uparrow\uparrow\rangle$ and
$|\!\!\downarrow\downarrow\rangle$.  For example,
\begin{equation}
    {\cal P}_{|\!\uparrow\uparrow\rangle} \, \rho_{AB}(t) \,
    P_{|\!\uparrow\uparrow\rangle} =
    \varkappa_{|\!\uparrow\uparrow\rangle}(t)
    \rho_{|\!\uparrow\uparrow\rangle}(t) ~,
\end{equation}
where ${\cal P}_{|\!\uparrow\uparrow\rangle} =
|\!\uparrow\uparrow\rangle \, \langle \!  \, \uparrow\uparrow |$ is
the projector onto state $|\!\uparrow\uparrow\rangle$, and
\begin{equation}
    \varkappa_{|\!\uparrow\uparrow\rangle}(t) = {\mathrm{Tr}} ( {\cal
    P}_{|\!\uparrow\uparrow\rangle} \, \rho_{AB}(t) \, {\cal
    P}_{|\!\uparrow\uparrow\rangle}) ~.
\end{equation}
After some algebra, we find that the correlation coefficients are
given in terms of the parameters of the density matrix in
Eq.~(\ref{Eq:4}) by
\begin{eqnarray}
     && \varkappa_{\frac{|\!\uparrow\downarrow\rangle +
     |\!\downarrow\uparrow\rangle}{\sqrt{2}}} = \frac{1}{4}\left[1 +
     {\bf n}_{A,x} {\bf n}_{B,x}+ {\bf n}_{A,y} {\bf n}_{B,y}- {\bf
     n}_{A,z} {\bf n}_{B,z}+ C_{xx}+C_{yy}-C_{zz} \right] ~,
     \label{Eq:G1} \\
    && \varkappa_{|\!\downarrow\downarrow\rangle} = \frac{1}{4}\left[1 -
    {\bf n}_{A,z} -{\bf n}_{B,z}+ {\bf n}_{A,z} {\bf n}_{B,z}+ C_{zz}
    \right] ~, \label{Eq:G2} \\
    && \varkappa_{|\!\uparrow\uparrow\rangle} = \frac{1}{4}\left[1 + {\bf
    n}_{A,z} +{\bf n}_{B,z}+ {\bf n}_{A,z} {\bf n}_{B,z}+C_{zz}
    \right] ~.
    \label{Eq:G3}
\end{eqnarray}
When the atoms are not correlated, the $C_{ij}$ terms vanish in these
formulas, and the correlation coefficients are given slowly in terms
of the expectation values $\{{\bf n}_{J,i}\}$ of the ``spins''.

In order to use these results to calculate the clock shifts for atomic
clocks using such atoms as ${}^{133}$Cs and ${}^{87}$Rb and
${}^{88}$Sr, three $s$-wave scattering lengths, $a_{s,gg}$, $a_{s,ee}$
and $a_{s,ge}$ would are required.  Hopefully, improved experimental
values for these $s$-wave scattering lengths, perhaps in conjunction
with theoretical calculations, will allow comparison of theory with
collisional clock shifts for bosonic atomic clocks.

\section{Summary and Conclusion}  \label{Summary_conclusion}

In summary, we developed a formalism that can be used to calculate
atomic clock shifts for arbitrary mixed states of an atomic gas.  We
applied the formalism to the fermionic $^{87}$Sr atomic clock reported
in Refs.~\cite{Campbell_09, BlattYe_09}.  When the two fermions that
collide in the gas are in a pure antisymmetric internal state,
$|\Psi_{AS} \rangle = \frac{1}{\sqrt{2}}(|\psi_{1} \rangle |\psi_{2}
\rangle - |\psi_{2} \rangle |\psi_{1} \rangle)$, the formalism reduces
to the one presented in Ref.~\cite{Campbell_09}.  For sufficiently
high temperatures (e.g., $T = 1$ or 3 $\mu$K), the gas may be closer
to an uncorrelated gas, and we can treat this case, as well as the
case of an arbitrary initial correlated state by projecting out the
two-body density matrix and using it to describe the collision process
of the atoms in the atomic clock.  We saw in Sec.~\ref{Clock_Calc}
that, for the experimental conditions described in
Refs.~\cite{Campbell_09,BlattYe_09}, when the internal state of
colliding pairs of atoms in the gas is taken as product mixed state
rather than an antisymmetric pure state, the collisional cloch shift
is somewhat reduced, but the reduction is small.  Moreover, we showed
how the formalism can be used to describe a bose gas; then, all the
$s$-wave scattering lengths ($a_{s,gg}, a_{s,ee}, a_{s,eg}$) are
required to calculate the collisional clock shift \cite{Band_00}.

The formalism developed here can also be applied to atomic clocks that
use the Ramsey fringe double resonance technique \cite{Vanier_Lemonde,
Band_00}.  Moreover it can be used for clocks that have a strongly
correlated initial state, since arbitrary two-particle density
matrices can be modeled using the form in Eq.~(\ref{Eq:4}).
Furthermore, even if the initial gas is a highly correlated many-body
state, the two-body density matrix can be projected out of the
many-body density matrix.  The formalism might also be of utility for
treating magnetometry wherein the collisions affect the
spin-relaxation time depolarization via depolarization.

Another possible application of the density matrix methods introduced
here is to calculation blackbody radiation (BBR) shifts.  BBR shifts
in atomic clocks are caused by perturbation of the atomic energy
levels of the ground and excited clock levels by the oscillating
thermal radiation.  Since we can carry out calculations with arbitrary
density matrix (mixed) states, our method could be used to calculate
blackbody radiation (BBR) effects, which can populate both ground and
excited clock states of an atomic clock operating at microwave
frequencies, if the clock operates at room temperature.  Moreover,
both ground and excited atomic levels of an optical frequency clock
transition are perturbed by BBR, e.g., for the ${}^1S_0 \, \to
{}^3$P$_0$ clock transition of strontium, both clock levels experience
a BBR shift \cite{Porsev_06}.  The overall BBR correction is the
difference of the BBR shifts for the two levels.  If there is a small
residual magnetic field present, the BBR could shift the energy of
ground hyperfine levels due to coupling between the various $F, M_F$
levels of the ground state.  Such shifts would be in addition to
shifts due to coupling of the ground and excited clock states to other
states.  A density matrix method of the type introduced here could be
used to deal with the former type of clock shift.

\begin{acknowledgments}
Useful discussions with Professor Jun Ye are gratefully acknowledged.
This work was supported in part by grants from the U.S.-Israel
Binational Science Foundation (No.~2006212), the Israel Science
Foundation (No.~29/07), and the James Franck German-Israel Binational
Program.
\end{acknowledgments}

\clearpage

\end{document}